\documentclass[onecolumn,preprintnumbers,amsmath,amssymb]{revtex4}

\usepackage{graphicx}
\usepackage{dcolumn}
\usepackage{bm}

\begin{document}

\title{Fast quantum dot single photon source triggered at telecommunications wavelength}

\author{Kelley Rivoire$^1$}
 \email{krivoire@stanford.edu}
\author{Sonia Buckley$^1$}
\author{Arka Majumdar$^1$}
\author{Hyochul Kim$^2$}
\author{Pierre Petroff$^2$}
\author{Jelena Vu\v{c}kovi\'{c}$^1$}%

\affiliation{%
$^1$E. L. Ginzton Laboratory, Stanford University, Stanford, CA\\ $^2$Department of Physics, University of California Santa Barbara, Santa Barbara, CA}%



\begin{abstract}

We demonstrate a quantum dot single photon source at 900 nm triggered at 300 MHz by a continuous wave telecommunications wavelength laser followed by an electro-optic modulator. The quantum dot is excited by on-chip-generated second harmonic radiation, resonantly enhanced by a GaAs photonic crystal cavity surrounding the InAs quantum dot. Our result suggests a path toward the realization of telecommunications-wavelength-compatible quantum dot single photon sources with speeds exceeding 1 GHz.

\end{abstract}
\maketitle
Single quantum emitters such as nitrogen vacancy centers \cite{beveratos}, molecules \cite{moerner}, semiconductor quantum dots \cite{michler}, and atoms \cite{rempe} emit antibunched light suitable for communications requiring single photons\cite{gisin,milburn}. Among these systems, semiconductor quantum dots have the highest emission rates and can be most easily integrated with semiconductor technology, including microcavities with high quality factor, small volume, and directional emission that increase the emission rate, efficiency, and indistinguishability of the generated single photons.\cite{santori, michler}

However, the generation rate of demonstrated optically triggered quantum dot single photon sources has been limited by excitation (Ti:Sapphire) lasers to around 80 MHz \cite{michler}. Electrical excitation \cite{stevenson, rakher} can circumvent this; however, resonant optical excitation \cite{santori} improves the indistinguishability of output photons, and many desirable microcavity structures such as photonic crystal cavities have geometries that are challenging to pump electrically \cite{bryan}. Furthermore, while telecommunications wavelengths are desirable for transporting photons over long distances, excitation and emission in many quantum dot materials systems, determined by material parameters, occurs at much shorter wavelengths.

Recently \cite{shg,sfg}, we demonstrated that photonic crystal cavities fabricated in III-V semiconductors with large $\chi^{(2)}$ nonlinearities can greatly enhance nonlinear frequency conversion efficiency, as a result of light recirculation inside an ultrasmall volume. Here, we apply a similar approach to excite a single InAs quantum dot (with transitions $\sim$900 nm) using a commercially available telecommunications wavelength ($\sim$1550 nm) laser that can serve as a trigger at GHz speeds when paired with a lithium niobate electro-optic modulator.

A scanning electron microscope image of the three-hole linear defect photonic crystal cavity\cite{noda} is shown in Fig. 1a. The structures are fabricated in a 164 nm thick GaAs membrane using e-beam lithography, dry etching, and HF wet etching of the sacrificial layer beneath the membrane, as described previously \cite{dirk_nature}. Quantum dots are grown by molecular beam epitaxy in the center of the membrane. Cavities are fabricated by e-beam lithography, dry etching, and HF release of the sacrificial layer. The cavity axis is oriented along a [011] crystal direction.The dot density is $<$10 dots/$\mu m^2$, and the quantum dot inhomogeneous broadening is around 40 nm, preventing multiple dots from having resonances at the same frequency.

Finite difference time domain simulations of the electric field components (TE-like resonance with electric field primarily in the plane of the photonic crystal slab) are shown in Fig. 1c. A reflectivity spectrum of the cavity showing a resonance at $\sim$1500 nm, measured with a broadband source in the cross-polarized configuration (as in our previous work \cite{APL_GaP}) to maximize signal to noise, is shown in Fig. 1b. A Lorentzian fit gives a cavity Q of 7,000. GaAs has a noncentrosymmetric cubic crystal lattice with 4$\overline{3}$m symmetry; the only non-zero elements of the bulk $\chi^{(2)}$ tensor have $i\neq j\neq k$. Since normally incident light couples to the TE-like photonic crystal cavity mode (Fig. 1c, with dominant $E_x$ and $E_y$ in-plane field components), the generated nonlinear polarization (and accordingly second harmonic, as described in \cite{shg}) is $\hat{z}$-polarized, i.e., transverse magnetic-like (TM-like) mode. For a photonic crystal, this corresponds to guided resonances in the TM-like air-band \cite{shg}. The second harmonic at 750 nm is used to excite the quantum dot above the GaAs band gap. Anisotropy of the quantum dot causes the dot to emit with TE polarization; at the quantum dot wavelength ($\sim$900 nm), the photonic crystal supports an optical (air-band) TE-like mode (Fig. 1d), weakly localized by the perturbation from removing and shifting holes near the cavity.

A tunable telecommunications wavelength continuous wave laser (Agilent 81989A, 1463 nm - 1577 nm) is coupled into the fundamental high-Q mode of the cavity. Photons upconverted in the cavity through second harmonic generation excite the quantum dot above band; the emitted single photons from the dot are spectrally filtered and sent to a spectrometer or Hanbury Brown-Twiss (HBT) setup for photon statistics analysis. The telecom wavelength laser can also be modulated for triggered single photon source operation. For direct recombination lifetime measurements of the quantum dot, we alternatively use an ultrashort pulse (3 ps) Ti:Sapphire laser to excite the dot and a streak camera for detection.

Continuous wave (CW) excitation is shown in Figure 2. Fig. 2a shows the photoluminescence spectrum measured when the continuous wave tunable laser, on resonance with the cavity mode (at $\sim$1500 nm), is normally incident on the structure. Spectral filters are used to create an effective bandpass filter ($\sim$5 nm FWHM) centered at 898 nm. Based on photon correlation measurements (Fig. 2b), it is likely that the two strongest emission lines (red box, Fig. 2a), both of which show linear dependence on pump power in the regime of low optical power, come from the same initial state of a single quantum dot.

To determine the maximum speed at which the system can be modulated, we perform an independent experiment directly measuring the dot lifetime with a streak camera (2 ps timing resolution), using a Ti:Sapphire laser at 750 nm with 80 MHz repetition rate and 3 ps pulses to excite the dot (Fig. 2a, inset). To accumulate sufficient counts, data from peaks shown in the red box in Fig. 2a were summed. The data were fit to a monoexponential decay with time constant 2.4$\pm$0.1 ns (where given error is one standard deviation).

 To verify that the upconverted light from the CW tunable laser could be used to efficiently excite a single quantum dot, we perform a photon correlation measurement of the signal collected at $\sim$900 nm (QD emission wavelength) using the HBT setup. Figure 2b shows the histogram of coincidence counts as a function of time delay $\tau$ between the two detectors. The data were fit to the unnormalized correlation function $G^{(2)}(\tau)=<I(t+\tau)I(t)>=A[1-(1-g^{(2)}(0))e^{-|\tau|/\tau_0}]$ \cite{arakawa}, where $A$, $\tau_0$ and $g^{(2)}(0)$ are fitting parameters, and $1/\tau_0=\Gamma+r_p$ where $\Gamma$ is the spontaneous emission lifetime and $r_p$ is the pump rate. The antibunching in coincidence counts at zero time delay g$^{(2)}(0)$=0.43$\pm$0.04, normalized such that g$^{(2)}(\infty)$=1, indicates emission primarily from the same initial state of a single quantum dot. Background subtraction for a signal to noise ratio of $S/N\approx10$ gives g$^{(2)}(0)$=0.31$\pm$0.05. An exponential fit to the central antibunching dip gives a decay rate of 2.3$\pm$0.2 ns; by comparison with independent lifetime measurements, this indicates that the dot is pumped in the low power regime, far below saturation.

To realize a triggered single photon source, the telecom tunable laser is modulated by a gigahertz electro-optic modulator (JDSU), which is driven by a pulse pattern generator (Anritsu 1800A) that receives an external clock signal from a synthesizer sweeper (Agilent). Photon correlation measurements were performed with the same HBT setup as in the CW experiment (Fig. 2). The repetition rate of the source was varied between 100 and 300 MHz, with duty cycle 20-50\% (minimum duty cycle for which the experiment could be performed was limited by background from the modulator during the off portion of the cycle; the finite excitation duration likely increases g$^{(2)}(0)$), as a result of the quantum dot re-excitation. Fig. 3a shows the measured photon statistics for a repetition rate of 100 MHz. The peaks were fit to exponentials with a single decay rate, and the ratio of the areas of the fitted peaks was used to determine g$^{(2)}(0)$=0.49$\pm$0.07 (g$^{(2)}(0)$=0.38$\pm$0.08 with background subtraction, indicating multi-photon probability suppression to 38\% relative to an attenuated laser of the same power). There is also a reduction in g$^{(2)}$ for peaks immediately adjacent to the central peak; this is likely a result of memory effects in the quantum dot \cite{santorisubmus}. Fig. 3b shows photon correlation measurements for repetition rate of 300 MHz; g$^{(2)}(0)$=0.40$\pm$0.09 (g$^{(2)}(0)$=0.3$\pm$0.1 with background subtraction). At 300 MHz, coincidences from consecutive pulses begin to overlap, as a result of the excitation repetition period approaching the quantum dot lifetime.

In conclusion, we have demonstrated an optically triggered single photon source operating at up to 300 MHz, excited with a telecommunications wavelength ($\sim$1500 nm) laser followed by an electro-optic modulator. In our system, the rate of single photons is limited primarily by the spontaneous emission rate of the quantum dot. A second cavity resonance at the frequency of the dot would increase the spontaneous emission rate of the dot via the Purcell effect\cite{jelena_apl_2003}, and enable the realization of a significantly faster (e.g. $>$10 GHz) triggered single photon source. Photons emitted from the quantum dot could also be frequency converted back to telecommunications wavelengths via difference frequency generation using the nonlinearity of the surrounding GaAs material \cite{mccutcheon} providing a full interface between the $\sim$900 nm quantum node and the optical fiber network.

This work was supported by the Army Research Office (Grant W911NF-08-1-0399), the National Science Foundation
(NSF Grant DMR-0757112), and the Air Force Office of Scientific Research (Agreement No: FA9550-09-1-0704). K.R., S.B., and A.M. are supported by Stanford Graduate Fellowships; S.B. is also supported by the NSF GRFP. The work was performed in part at the Stanford Nanofabrication Facility of NNIN. We thank Jason Pelc, Qiang Zhang, Seth Lloyd and Vicky Wen for lending equipment.


\begin{thebibliography}{19}
\expandafter\ifx\csname natexlab\endcsname\relax\def\natexlab#1{#1}\fi
\expandafter\ifx\csname bibnamefont\endcsname\relax
  \def\bibnamefont#1{#1}\fi
\expandafter\ifx\csname bibfnamefont\endcsname\relax
  \def\bibfnamefont#1{#1}\fi
\expandafter\ifx\csname citenamefont\endcsname\relax
  \def\citenamefont#1{#1}\fi
\expandafter\ifx\csname url\endcsname\relax
  \def\url#1{\texttt{#1}}\fi
\expandafter\ifx\csname urlprefix\endcsname\relax\def\urlprefix{URL }\fi
\providecommand{\bibinfo}[2]{#2}
\providecommand{\eprint}[2][]{\url{#2}}

\bibitem[{\citenamefont{Beveratos et~al.}(2001)\citenamefont{Beveratos, Brouri,
  Gacoin, Poizat, and Grangier}}]{beveratos}
\bibinfo{author}{\bibfnamefont{A.}~\bibnamefont{Beveratos}},
  \bibinfo{author}{\bibfnamefont{R.}~\bibnamefont{Brouri}},
  \bibinfo{author}{\bibfnamefont{T.}~\bibnamefont{Gacoin}},
  \bibinfo{author}{\bibfnamefont{J.-P.} \bibnamefont{Poizat}},
  \bibnamefont{and} \bibinfo{author}{\bibfnamefont{P.}~\bibnamefont{Grangier}},
  \bibinfo{journal}{Phys. Rev. A} \textbf{\bibinfo{volume}{64}},
  \bibinfo{pages}{061802(R)} (\bibinfo{year}{2001}).

\bibitem[{\citenamefont{Lounis and Moerner}(2000)}]{moerner}
\bibinfo{author}{\bibfnamefont{B.}~\bibnamefont{Lounis}} \bibnamefont{and}
  \bibinfo{author}{\bibfnamefont{W.}~\bibnamefont{Moerner}},
  \bibinfo{journal}{Nature} \textbf{\bibinfo{volume}{407}},
  \bibinfo{pages}{491} (\bibinfo{year}{2000}).

\bibitem[{\citenamefont{Michler et~al.}(2000)\citenamefont{Michler, Kiraz,
  Becher, Schoenfeld, Petroff, Zhang, Hu, and Imamoglu}}]{michler}
\bibinfo{author}{\bibfnamefont{P.}~\bibnamefont{Michler}},
  \bibinfo{author}{\bibfnamefont{A.}~\bibnamefont{Kiraz}},
  \bibinfo{author}{\bibfnamefont{C.}~\bibnamefont{Becher}},
  \bibinfo{author}{\bibfnamefont{W.}~\bibnamefont{Schoenfeld}},
  \bibinfo{author}{\bibfnamefont{P.}~\bibnamefont{Petroff}},
  \bibinfo{author}{\bibfnamefont{L.}~\bibnamefont{Zhang}},
  \bibinfo{author}{\bibfnamefont{E.}~\bibnamefont{Hu}}, \bibnamefont{and}
  \bibinfo{author}{\bibfnamefont{A.}~\bibnamefont{Imamoglu}},
  \bibinfo{journal}{Science} \textbf{\bibinfo{volume}{290}},
  \bibinfo{pages}{2282} (\bibinfo{year}{2000}).

\bibitem[{\citenamefont{Kohn et~al.}(2002)\citenamefont{Kohn, Hennrich, and
  Rempe}}]{rempe}
\bibinfo{author}{\bibfnamefont{A.}~\bibnamefont{Kohn}},
  \bibinfo{author}{\bibfnamefont{M.}~\bibnamefont{Hennrich}}, \bibnamefont{and}
  \bibinfo{author}{\bibfnamefont{G.}~\bibnamefont{Rempe}},
  \bibinfo{journal}{Phys. Rev. Lett.} \textbf{\bibinfo{volume}{89}},
  \bibinfo{pages}{067901} (\bibinfo{year}{2002}).

\bibitem[{\citenamefont{Gisin et~al.}(2002)\citenamefont{Gisin, Ribordy,
  Tittel, and Zbinden}}]{gisin}
\bibinfo{author}{\bibfnamefont{N.}~\bibnamefont{Gisin}},
  \bibinfo{author}{\bibfnamefont{G.}~\bibnamefont{Ribordy}},
  \bibinfo{author}{\bibfnamefont{W.}~\bibnamefont{Tittel}}, \bibnamefont{and}
  \bibinfo{author}{\bibfnamefont{H.}~\bibnamefont{Zbinden}},
  \bibinfo{journal}{Rev. Mod. Phys.} \textbf{\bibinfo{volume}{74}},
  \bibinfo{pages}{145} (\bibinfo{year}{2002}).

\bibitem[{\citenamefont{Knill et~al.}(2001)\citenamefont{Knill, Laflamme, and
  Milburn}}]{milburn}
\bibinfo{author}{\bibfnamefont{E.}~\bibnamefont{Knill}},
  \bibinfo{author}{\bibfnamefont{R.}~\bibnamefont{Laflamme}}, \bibnamefont{and}
  \bibinfo{author}{\bibfnamefont{G.}~\bibnamefont{Milburn}},
  \bibinfo{journal}{Nature} \textbf{\bibinfo{volume}{409}}, \bibinfo{pages}{46}
  (\bibinfo{year}{2001}).

\bibitem[{\citenamefont{Santori et~al.}(2002)\citenamefont{Santori, Fattal,
  Vu\v{c}kovi\'{c}, Solomon, and Yamamoto}}]{santori}
\bibinfo{author}{\bibfnamefont{C.}~\bibnamefont{Santori}},
  \bibinfo{author}{\bibfnamefont{D.}~\bibnamefont{Fattal}},
  \bibinfo{author}{\bibfnamefont{J.}~\bibnamefont{Vu\v{c}kovi\'{c}}},
  \bibinfo{author}{\bibfnamefont{G.}~\bibnamefont{Solomon}}, \bibnamefont{and}
  \bibinfo{author}{\bibfnamefont{Y.}~\bibnamefont{Yamamoto}},
  \bibinfo{journal}{Nature} \textbf{\bibinfo{volume}{419}},
  \bibinfo{pages}{494} (\bibinfo{year}{2002}).

\bibitem[{\citenamefont{Yuan et~al.}(2002)\citenamefont{Yuan, Kardynal,
  Stevenson, Shields, Lobo, Cooper, Beattie, Ritchie, and Pepper}}]{stevenson}
\bibinfo{author}{\bibfnamefont{Z.}~\bibnamefont{Yuan}},
  \bibinfo{author}{\bibfnamefont{B.}~\bibnamefont{Kardynal}},
  \bibinfo{author}{\bibfnamefont{R.}~\bibnamefont{Stevenson}},
  \bibinfo{author}{\bibfnamefont{A.}~\bibnamefont{Shields}},
  \bibinfo{author}{\bibfnamefont{C.}~\bibnamefont{Lobo}},
  \bibinfo{author}{\bibfnamefont{K.}~\bibnamefont{Cooper}},
  \bibinfo{author}{\bibfnamefont{N.}~\bibnamefont{Beattie}},
  \bibinfo{author}{\bibfnamefont{D.}~\bibnamefont{Ritchie}}, \bibnamefont{and}
  \bibinfo{author}{\bibfnamefont{M.}~\bibnamefont{Pepper}},
  \bibinfo{journal}{Science} \textbf{\bibinfo{volume}{295}},
  \bibinfo{pages}{102} (\bibinfo{year}{2002}).

\bibitem[{\citenamefont{Rakher et~al.}(2008)\citenamefont{Rakher, Stoltz,
  Coldren, Petroff, and Bouwmeester}}]{rakher}
\bibinfo{author}{\bibfnamefont{M.~T.} \bibnamefont{Rakher}},
  \bibinfo{author}{\bibfnamefont{N.~G.} \bibnamefont{Stoltz}},
  \bibinfo{author}{\bibfnamefont{L.~A.} \bibnamefont{Coldren}},
  \bibinfo{author}{\bibfnamefont{P.~M.} \bibnamefont{Petroff}},
  \bibnamefont{and}
  \bibinfo{author}{\bibfnamefont{D.}~\bibnamefont{Bouwmeester}},
  \bibinfo{journal}{Appl. Phys. Lett.} \textbf{\bibinfo{volume}{93}},
  \bibinfo{eid}{091118} (pages~\bibinfo{numpages}{3}) (\bibinfo{year}{2008}).

\bibitem[{\citenamefont{Ellis et~al.}(2010)\citenamefont{Ellis, Sarmiento,
  Mayer, Zhang, Harris, Haller, and Vuckovic}}]{bryan}
\bibinfo{author}{\bibfnamefont{B.}~\bibnamefont{Ellis}},
  \bibinfo{author}{\bibfnamefont{T.}~\bibnamefont{Sarmiento}},
  \bibinfo{author}{\bibfnamefont{M.}~\bibnamefont{Mayer}},
  \bibinfo{author}{\bibfnamefont{B.}~\bibnamefont{Zhang}},
  \bibinfo{author}{\bibfnamefont{J.}~\bibnamefont{Harris}},
  \bibinfo{author}{\bibfnamefont{E.}~\bibnamefont{Haller}}, \bibnamefont{and}
  \bibinfo{author}{\bibfnamefont{J.}~\bibnamefont{Vuckovic}},
  \bibinfo{journal}{Appl. Phys. Lett.} \textbf{\bibinfo{volume}{96}},
  \bibinfo{pages}{181103} (\bibinfo{year}{2010}).

\bibitem[{\citenamefont{Rivoire et~al.}(2009)\citenamefont{Rivoire, Lin,
  Hatami, Masselink, and Vu\v{c}kovi\'{c}}}]{shg}
\bibinfo{author}{\bibfnamefont{K.}~\bibnamefont{Rivoire}},
  \bibinfo{author}{\bibfnamefont{Z.}~\bibnamefont{Lin}},
  \bibinfo{author}{\bibfnamefont{F.}~\bibnamefont{Hatami}},
  \bibinfo{author}{\bibfnamefont{W.~T.} \bibnamefont{Masselink}},
  \bibnamefont{and}
  \bibinfo{author}{\bibfnamefont{J.}~\bibnamefont{Vu\v{c}kovi\'{c}}},
  \bibinfo{journal}{Opt. Express} \textbf{\bibinfo{volume}{17}},
  \bibinfo{pages}{22609} (\bibinfo{year}{2009}).

\bibitem[{\citenamefont{Rivoire et~al.}(2010)\citenamefont{Rivoire, Lin,
  Hatami, and Vu\v{c}kovi\'{c}}}]{sfg}
\bibinfo{author}{\bibfnamefont{K.}~\bibnamefont{Rivoire}},
  \bibinfo{author}{\bibfnamefont{Z.}~\bibnamefont{Lin}},
  \bibinfo{author}{\bibfnamefont{F.}~\bibnamefont{Hatami}}, \bibnamefont{and}
  \bibinfo{author}{\bibfnamefont{J.}~\bibnamefont{Vu\v{c}kovi\'{c}}},
  \bibinfo{journal}{Appl. Phys. Lett.} \textbf{\bibinfo{volume}{97}},
  \bibinfo{pages}{043103} (\bibinfo{year}{2010}).

\bibitem[{\citenamefont{Akahane et~al.}(2003)\citenamefont{Akahane, Asano,
  Song, and Noda}}]{noda}
\bibinfo{author}{\bibfnamefont{Y.}~\bibnamefont{Akahane}},
  \bibinfo{author}{\bibfnamefont{T.}~\bibnamefont{Asano}},
  \bibinfo{author}{\bibfnamefont{B.}~\bibnamefont{Song}}, \bibnamefont{and}
  \bibinfo{author}{\bibfnamefont{S.}~\bibnamefont{Noda}},
  \bibinfo{journal}{Nature} \textbf{\bibinfo{volume}{425}},
  \bibinfo{pages}{944} (\bibinfo{year}{2003}).

\bibitem[{\citenamefont{Englund et~al.}(2007)\citenamefont{Englund, Faraon,
  Fushman, Stoltz, Petroff, and Vuckovic}}]{dirk_nature}
\bibinfo{author}{\bibfnamefont{D.}~\bibnamefont{Englund}},
  \bibinfo{author}{\bibfnamefont{A.}~\bibnamefont{Faraon}},
  \bibinfo{author}{\bibfnamefont{I.}~\bibnamefont{Fushman}},
  \bibinfo{author}{\bibfnamefont{N.}~\bibnamefont{Stoltz}},
  \bibinfo{author}{\bibfnamefont{P.}~\bibnamefont{Petroff}}, \bibnamefont{and}
  \bibinfo{author}{\bibfnamefont{J.}~\bibnamefont{Vuckovic}},
  \bibinfo{journal}{Nature.} \textbf{\bibinfo{volume}{450}}
  (\bibinfo{year}{2007}).

\bibitem[{\citenamefont{Rivoire et~al.}(2008)\citenamefont{Rivoire, Faraon, and
  Vuckovic}}]{APL_GaP}
\bibinfo{author}{\bibfnamefont{K.}~\bibnamefont{Rivoire}},
  \bibinfo{author}{\bibfnamefont{A.}~\bibnamefont{Faraon}}, \bibnamefont{and}
  \bibinfo{author}{\bibfnamefont{J.}~\bibnamefont{Vuckovic}},
  \bibinfo{journal}{Appl. Phys. Lett.} \textbf{\bibinfo{volume}{93}}
  (\bibinfo{year}{2008}).

\bibitem[{\citenamefont{Kako et~al.}(2006)\citenamefont{Kako, Santori, Hoshino,
  Gotzinger, Yamamoto, and Arakawa}}]{arakawa}
\bibinfo{author}{\bibfnamefont{S.}~\bibnamefont{Kako}},
  \bibinfo{author}{\bibfnamefont{C.}~\bibnamefont{Santori}},
  \bibinfo{author}{\bibfnamefont{K.}~\bibnamefont{Hoshino}},
  \bibinfo{author}{\bibfnamefont{S.}~\bibnamefont{Gotzinger}},
  \bibinfo{author}{\bibfnamefont{Y.}~\bibnamefont{Yamamoto}}, \bibnamefont{and}
  \bibinfo{author}{\bibfnamefont{Y.}~\bibnamefont{Arakawa}},
  \bibinfo{journal}{Nat. Mat.} \textbf{\bibinfo{volume}{5}},
  \bibinfo{pages}{887} (\bibinfo{year}{2006}).

\bibitem[{\citenamefont{Santori et~al.}(2004)\citenamefont{Santori, Fattal,
  Vu\v{c}kovi\'{c}, Solomon, Waks, and Yamamoto}}]{santorisubmus}
\bibinfo{author}{\bibfnamefont{C.}~\bibnamefont{Santori}},
  \bibinfo{author}{\bibfnamefont{D.}~\bibnamefont{Fattal}},
  \bibinfo{author}{\bibfnamefont{J.}~\bibnamefont{Vu\v{c}kovi\'{c}}},
  \bibinfo{author}{\bibfnamefont{G.}~\bibnamefont{Solomon}},
  \bibinfo{author}{\bibfnamefont{E.}~\bibnamefont{Waks}}, \bibnamefont{and}
  \bibinfo{author}{\bibfnamefont{Y.}~\bibnamefont{Yamamoto}},
  \bibinfo{journal}{Phys. Rev. B} \textbf{\bibinfo{volume}{69}},
  \bibinfo{pages}{205324} (\bibinfo{year}{2004}).

\bibitem[{\citenamefont{Vu\v{c}kovi\'{c}
  et~al.}(2003)\citenamefont{Vu\v{c}kovi\'{c}, Fattal, Santori, Solomon, and
  Yamamoto}}]{jelena_apl_2003}
\bibinfo{author}{\bibfnamefont{J.}~\bibnamefont{Vu\v{c}kovi\'{c}}},
  \bibinfo{author}{\bibfnamefont{D.}~\bibnamefont{Fattal}},
  \bibinfo{author}{\bibfnamefont{C.}~\bibnamefont{Santori}},
  \bibinfo{author}{\bibfnamefont{G.}~\bibnamefont{Solomon}}, \bibnamefont{and}
  \bibinfo{author}{\bibfnamefont{Y.}~\bibnamefont{Yamamoto}},
  \bibinfo{journal}{Appl. Phys. Lett.} \textbf{\bibinfo{volume}{82}},
  \bibinfo{pages}{3596} (\bibinfo{year}{2003}).

\bibitem[{\citenamefont{McCutcheon et~al.}(2009)\citenamefont{McCutcheon,
  Chang, Zhang, Lukin, and Loncar}}]{mccutcheon}
\bibinfo{author}{\bibfnamefont{M.~W.} \bibnamefont{McCutcheon}},
  \bibinfo{author}{\bibfnamefont{D.~E.} \bibnamefont{Chang}},
  \bibinfo{author}{\bibfnamefont{Y.}~\bibnamefont{Zhang}},
  \bibinfo{author}{\bibfnamefont{M.~D.} \bibnamefont{Lukin}}, \bibnamefont{and}
  \bibinfo{author}{\bibfnamefont{M.}~\bibnamefont{Loncar}},
  \bibinfo{journal}{Opt. Express} \textbf{\bibinfo{volume}{17}},
  \bibinfo{pages}{22689} (\bibinfo{year}{2009}).

\end{thebibliography}
%

 \begin{figure}[h]
\includegraphics{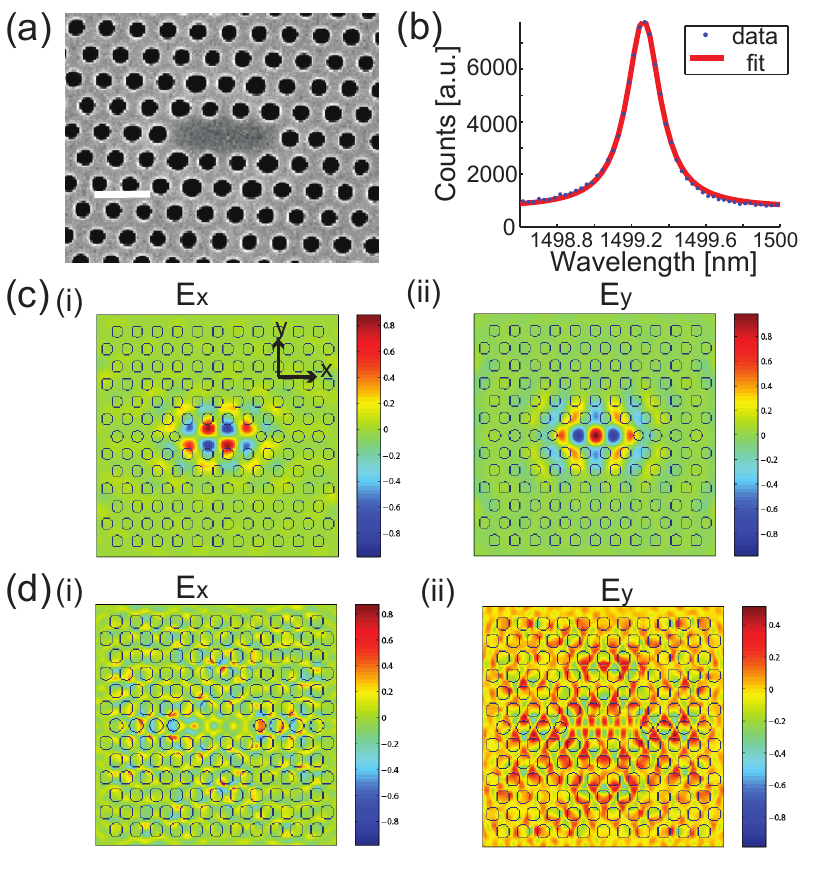}
\caption{\label{fig:pc2}
  Photonic crystal cavity design and fabrication. (a) SEM image of fabricated suspended membrane photonic crystal cavity. Scale bar indicates 1 $\mu$m. (b) Reflectivity measurement of fundamental cavity mode. Lorentzian fit gives Q of 7000. (c) (i) and (ii) Simulated electric field components for fundamental mode of L3 photonic crystal cavity, used for resonantly enhanced upconversion. (d) (i) and (ii) Electric field patterns for TE mode closest to emission frequency of quantum dot. The low Q ($<$100) mode is formed through weak confinement of an air band mode.}
\end{figure}

\begin{figure}[h]
\includegraphics{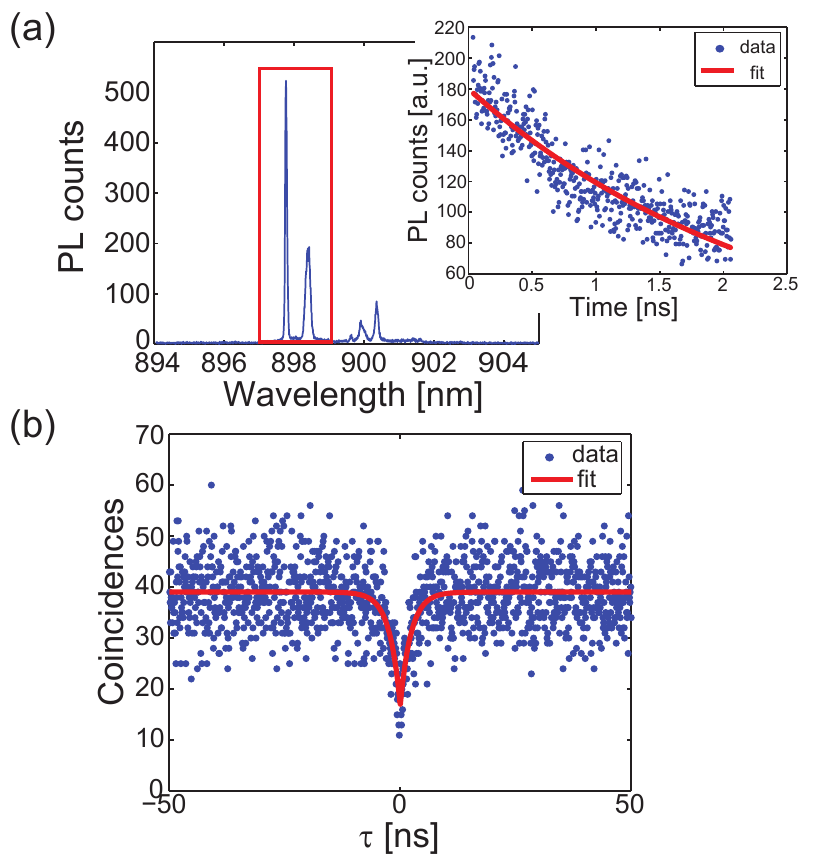}
\caption{\label{fig:pc}
 Characterization of single quantum dot excited with on-chip-upconverted 1500 nm laser. (a) Spectrum measured from CW second harmonic excitation of quantum dot. Inset: Streak camera measurement of quantum dot lifetime. Counts were summed over spectral window indicated by box. (b) Photon correlation measurement of quantum dot emission under frequency doubled CW 1550nm excitation. Fit gives g$^{(2)}$(0)=0.43$\pm$0.04.}
\end{figure}

\begin{figure}[h]
\includegraphics{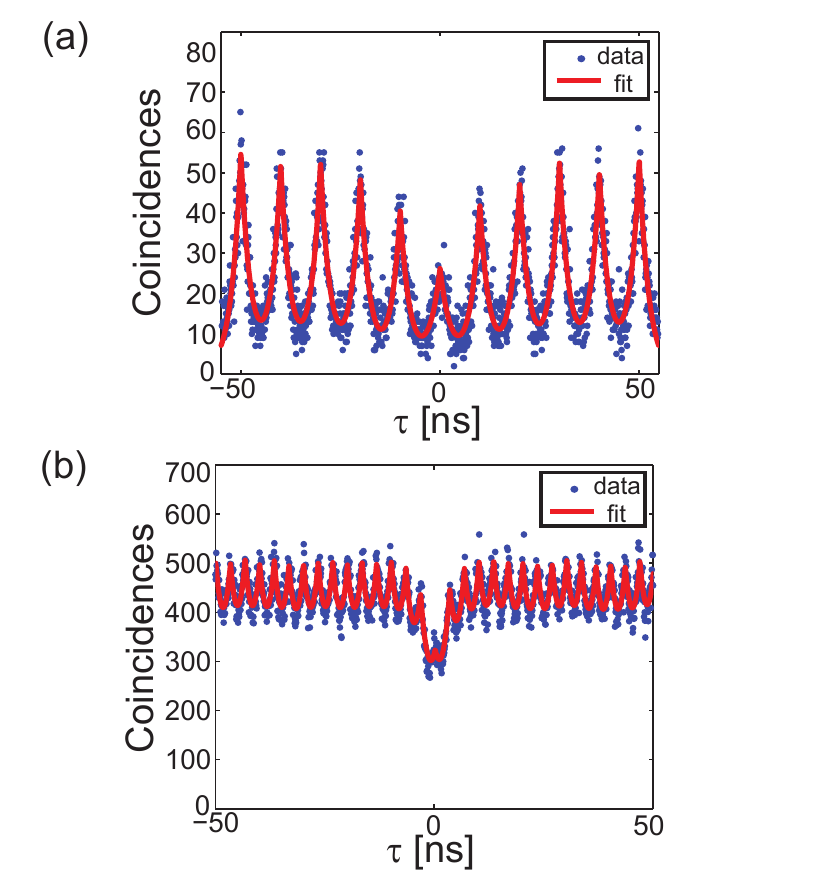}
\caption{\label{fig:pc2}
  Photon correlation measurement of quantum dot emission when triggered with frequency doubled telecom wavelength laser pulses from an externally modulated telecom wavelength laser. (a) Second order autocorrelation function measurement for 100 MHz repetition rate with duty
cycle 20\%. g$^{(2)}$(0) = 0.48 $<$ 0.5 indicates emission from a single quantum
dot. (b) Second order autocorrelation measurement for repetition rate 300 MHz
with duty cycle 50\% and g$^{(2)}$(0)=0.40.}
\end{figure}

\end{document}